\documentstyle[12pt]{article}

\def \c {\cite}
\def \ur {\uparrow}
\def \dr {\downarrow}

\begin{document}
\begin{titlepage}
\title {\bf Dihyperon in Chiral Colour Dielectric model }
\author{{\bf Sanjay K. Ghosh$^a${\thanks {email: phys@boseinst.ernet.in}},
and S. C. Phatak$^b${\thanks {email: phatak@iopb.ernet.in}}} \\
a) Department of Physics, Bose Institute, \\ 93/1, A. P. C. Road, Calcutta
700 009, INDIA \\
b) Institute of Physics, Bhubaneswar 751 005, INDIA}
\maketitle
\begin{center}
[PACS : 12.39.Ki, 12.39.Hk, 14.20.Pt] 
\end{center}
\begin{abstract}
The mass of dihyperon with spin, parity $J^{\pi}=0^{+}$ and isospin 
$I = 0$ is calculated in the framework of Chiral colour dielectric model.
The wave function of the dihyperon is expressed as a product of 
two colour-singlet baryon 
clusters. Thus the quark wave functions within the cluster are 
antisymmetric. Appropriate operators are then used to antisymmetrize 
inter-cluster quark wave functions. The radial part of the quark 
wavefunctions are obtained by solving the the quark
and dielectric field equations of motion obtained in  
the Colour dielectric model. The mass of the dihyperon is computed by 
including the colour magnetic energy as well as the energy due to meson 
interaction. The recoil correction to the dihyperon mass  is incorporated by 
Peierls-Yoccoz technique. We find that the mass of the dihyperon is smaller
than the $\Lambda-\Lambda$ threshold by over 100 MeV.  
The implications of our results on the present day relativistic heavy ion 
experiments is discussed.
\end{abstract}
\end{titlepage}
The possibility of the existence of a stable six quark dibaryon composed of two 
up (u), two down (d) and two strange (s) quarks confined in a single hadronic  
bag and having spin, parity $J^{\pi}=0^{+}$ and isospin $I = 0$ was first  
proposed by Jaffe in 1977 \c{jaf77}. This object is a  singlet of colour, 
flavour and spin and thus results in a maximally attractive colour magnetic 
interaction between the quarks. Jaffe's calculation predicted the  mass of 
dihyperon ($H_1$) to be
around 2150 MeV which is about 80 MeV less than the $\Lambda- \Lambda$ 
threshold.
Such a state would then be stable aganist strong decays into
three-quark baryons and can decay only by weak interaction into a pair of
baryons. Jaffe's calculation was performed in the MIT bag model. Later, this 
calculation was refined by including center-of-mass correction\c{liu82}, 
SU(3)-flavour symmetry breaking\c{rosen86},  surface energy term
in the bag model\c{aert84b}, coupling of pseudo-scalar meson octate
\c{mul83} etc. Calculations have also been
performed in non-relativistic potential model\c{oka83,silve87,wag95} 
and Skyrmion model\c{bala84}.
Production cross sections of dihyperons in various experiments have also been
estimeted\c{aert84a}. Most of these calculations predict the mass of $H_1$
very close to $\Lambda-\Lambda$ threshold. Some of these calculations predict
that  $H_1$ is stable against strong decays with mass below the 
$\Lambda - \Lambda$ threshold where as
the others predict an unstable $H_1$.
If the dihyperon mass is below the $\Lambda-\Lambda$ 
threshold, one expects that, unlike the deuteron, the dihyperon would be 
a state of six-quarks bound in a single bag and not a two-baryon state 
bound by meson exhange 
interaction. The reason for this is that in the deuteron 
the pion exchange interaction provides the bulk of the binding force where as
in the dihyperon case, one-pion exchange is not possible in $\Lambda-\Lambda$
channel and therefore the meson exchange contribution to the binding
is expected to be small.  Therefore the 
experimental determination of the dihyperon mass is expected to impose a strong 
constraint on the quark models used in hadron spectroscopy.

With these considerations, the experimental as well as theoretical investigations 
of dihyperons in particular and dibaryons in general 
is of great interest. As is well known, the  QCD is the theory of
strong interactions and in such a theory, six-quark colour-neutral objects 
are expected to exist. Whether these are stable against strong decays depends
on the details. Already a  number of QCD-inspired models have been employed to 
investigate the properties of baryons, the three-quark colour-neutral objects 
and generally these models are quite successful. 
The calculations of the properties of dihyperons and other dibaryons in these 
models and comparison of these with
the  experimental results is needed as these models are likely
to yield different results in the dibaryon sector. The experimental observation  
( or otherwise ) of the dihyperon will then be able to indicate 
which of these  models are better. 

The dihyperon, if stable, is likely to be produced in relativistic heavy ion 
collisions due to the abundant strangeness production in the hot and 
dense hadronic matter formed in the collision. 
For example, the calculations using a cascade code like
ARC \c{pang92} find that in a collision of Au ions with similarly heavy target
nuclei, more than 20 hyperons are expected to be produced in central 
collisions at the AGS energies. This implies that there is a large  probability
of $\Lambda-\Lambda$ coalescence leading to the formation of a dihyperon.
Further more, the formation of quark gluon
plasma with large baryon density and its subsequent decay is also expected 
to  enhance the strangeness production in the fragmentation region. This would
lead to an enhancement in the formation of dihyperons in the relativistic
heavy ion collisions.
So far, evidence for the existance of dihyperon or otherwise is rather scanty  
and inconclusive.
An isolated H candidate has been reported in bubble chamber 
experiments \c{shahba88}. Three H particle candidates have been observed 
in three different emulsion experiments \c{danysz63}.In an another 
experiment of heavy ion collision of Au + Pt, E886 collaboration has 
reported a null result for the search of H particle \c{rusek95}. On the 
other hand in a more recent heavy ion experiment\c{Long} a number of dihyperons 
seem to have been detected. This experiment seems
to give the dihyperon mass of about 2180 MeV and the lifetime of about 
$3.3 \times 10^{-10}$ sec. Of course, it  must noted that this result is 
not yet conclusive enough.

In the present work the mass of dihyperon has been calculated in the 
framework of chiral
colour dielectric (CCD) model. The  CCD model has been used earlier in 
baryon spectroscopy \c{9} and for the investigation of static properties of 
nucleons in nuclear medium\c{10}. These calculations have shown that the 
model is able to explain the static properties of light baryons very well. 
Furthermore, when applied to the quark matter calculation, the model yields an 
equation of state which is quite similar to the one obtained from lattice 
calculations for zero baryon chemical potential \c{11}. The  CCD model 
differs from the bag model in several aspects. First of all, in the CCD 
model, the confinement of quarks and gluons is achieved dynamically through 
the colour-dielectric field. In the bag model this is done by hand. Also the 
quark masses used in the CCD model are different from those used in the bag 
model. In the bag model, $u$ and $d$ masses are taken to be zero. The CCD 
model requires that these masses are nonzero. It has been found \c{9} that 
to fit baryon masses, the required $u$ and $d$ masses are $\sim
100MeV$. It might be noted that similar masses are used in relativistic 
quark models \c{12}. Thus, the values of quark masses in the CCD model
are closer to the constituent quark masses.
The main difference between the  present calculation and most of the earlier  
dihyperon calculations is the inclusion of the pseudoscalar meson coupling  
to the dihyperon state ( however see \c{mul83} and \c{wag95} which include
meson self energy ). The meson self-energy
corrections are expected to shift the masses of dibaryons by a few 
hundreds of MeV, just like the shifts produced  in the baryon masses. 
Therefore one needs to include these self energy contributions in a more realistic
calculation. Further more, the explicit
breaking of SU(3)-flavour symmetry, which arises from the difference between 
strange and u ( d ) quark masses is included in our calculation. 

The methodology adopted in the present work is similar to the one followed
in the baryon spectroscopy calculations\c{9}, or for that matter the one 
followed in the cloudy bag model calculations\c{12a}. Thus we first compute the quark 
wavefunctions in the mean field approximation by solving coupled quark 
and dielectric field equations obtained in the CCD model. These equations are
\begin{eqnarray}
(\vec \alpha \cdot \vec p - m - \frac{m_0}{\chi(r)} ) \Psi(r) &=& E \Psi(r) \\
\chi''(r) + \frac{2}{r} \chi '(r) - \frac{\partial V(\chi)}{\partial \chi}
+ \frac{m_0}{\sigma^2_v \chi^2(r)} <\overline \Psi(r) \Psi(r)> &=& 0.
\end{eqnarray}
Here $\Psi(r)$ is the four-component Dirac spinor, $\chi(r)$ is the colour 
dielectric field, $m_0$ is the u and d quark masses, $m$ is the mass 
difference
between s and u quarks, $\sigma^2_v = 2 \alpha B / m_{GB}^2$, $m_{GB}$ is the 
mass of the dielectric field and $V(\chi)$ is the self interaction of the 
dielectric field; 
$$ V(\chi) = B ( \alpha \chi^2 - 2(\alpha-2)\chi^3 + (\alpha-3)\chi^4).$$
In the mean field equations above,
 we have assumed  a spherically symmetric and time independent dielectric
field generated by the quarks present in the $s_{1/2}$ orbital.
(for the details of the CDM Lagrangian, field equations, mean field solutions 
 etc  the reader 
is referred to ref\c{9}.)
The wavefunction of a six-quark cluster is then constructed 
as a product of a six-quark space wavefunction and a
spin-flavour-colour wavefunction. Since all the six quarks 
are in $s_{1/2}$ orbital, the space wavefunction is symmetric. We find it 
convenient to 
express the spin-flavour-colour wavefunction as a product of 
two colour-neutral three-quark wavefunctions and then antisymmetrize the 
wavefunction with respect to the  exchange of quarks between the two 
clusters:
\begin{eqnarray}
|c_{1}c_{2}>_{f,s,c}&=&{\cal P}\sum_{1,2} \alpha_{1,2}|c_{1}>_{f} |c_{1}>_{s} 
|c_{1}>_c
\times |c_{2}>_{f} |c_{2}>_{s} |c_2>_c
\label{clust}
\end{eqnarray}
where the subscripts f, s, c denote the flavour, spin and colour wave 
functions of three-quark clusters and 
$c_{1}$ and $c_{2}$ denote the first and second cluster respectively. The 
colour wavefunction $|c_{i}>_c = \sum_{l,m,n} \epsilon_{l,m,n} |l>|m>|n>$
is antisymmetric ( colour-singlet ) with respect to exchange and that is ensured by the 
Levi-Civita symbol $\epsilon_{l,m,n}$.
$\sum_{1,2}$ includes the summation over possible spins,
isospins and hypercharges of the clusters 1 and 2 so as to give a dibaryon
state of definite spin, parity, isospin and strangeness.  
The permutation operator 
\begin{eqnarray}
{\cal P}&=&{1\over {\sqrt{8}}} [1 + {{\cal S}^{c}}_{14} {{\cal A}^{fs}}_{14}]
[ 1 + {{\cal S}^{c}}_{25} {{\cal A}^{fs}}_{25}] 
[ 1 + {{\cal S}^{c}}_{36} {{\cal A}^{fs}}_{36}]
\label{symm}
\end{eqnarray}
is required for proper antisymmetrization of quark wavefunctions between two
clusters. Note that since the colour wavefunction of a cluster is a colour 
singlet, we need to symmetrize intercluster colour wavefunction and therefore
antisymmetrize the spin-flavour wavefunction. Also, the spin and flavour 
cluster wavefunctions are symmetric and therefore are simply  the octate 
and decuplet baryon wavefunctions. The state thus constructed is a bare
dibaryon state and corrections due to gluon and pseudoscalar meson interactions
need to be included.

The dihyperon state we want to consider in this work is a colour- 
flavour- and spin-singlet state. In terms of the cluster wavefunction 
described above, the spin-flavour-colour wavefunction of the dihyperon is,
\begin{eqnarray}
|H_1>&=& {1\over 4}{\cal P}\{|p \Xi^{-}> + |\Xi^{-} p> - |n \Xi^{0}> - |\Xi^{0} n>
- |\Sigma^{+} \Sigma^{-}> - |\Sigma^{-} \Sigma^{+}> \nonumber \\
&+& |\Sigma^{0} \Sigma^{0}>
+ |\Lambda \Lambda> \} \{\ur \dr - \dr \ur\} |C_{1}>_c |C_2>_c
\label{hwav}
\end{eqnarray}
where the first term on the right hand side of eq. (\ref{hwav}) 
consists of a combination of
baryon octet flavour wavefunctions and the second bracket is  the antisymmetric
(two baryon) spin wave function. Note that the baryon wavefunctions themselves 
consist of the product of SU(3) colour and SU(6) flavour-spin wave functions 
of quarks. The fact that the wavefunction $|H_1>$ is a singlet of colour and  
spin is obvious. One can convince oneself  that it is a flavour singlet 
as well by showing  
that the operation of ( quark ) isospin, V-spin and U-spin raising and lowering 
operators on $|H_1>$ gives zero.

Let us now consider the mass of the dihyperon. In the cloudy bag model approach, 
the mass is given by
\begin{eqnarray}
M_{H_1} &=& M_{bare} + M_{c} + M_{meson}
\end{eqnarray}
where $M_{bare}$ is the contribution to the mass from the quarks and the 
dielectric field, $M_c$ is the contribution due to colour magnetic energy
and $M_{meson}$ is the meson self-energy correction. The bare mass $M_{bare}$
includes the energies of the quarks and the dielectric field. The
energy due to the spurious center of mass motion is removed by using the 
Peierls-Yoccoz projection technique\c{PY} in our calculation ( see ref\c{9}
for details). The colour magnetic energy $M_c$ is given by
\begin{eqnarray}
M_c = \frac{1}{2} \sum_{i < j} \int d^3r \vec j^a_i(\vec r) \cdot 
\vec A^a_j(\vec r)
\end{eqnarray}
where $\vec A^a_j(\vec r)$ is the ( colour ) vector potential generated by 
$j^{th}$ quark and $\vec j^a_i(\vec r)$ is the colour current of $i^{th}$
quark. The vector potential $\vec A$ is obtained by using Green's function
technique\c{9,Willets}. Thus, we have
\begin{eqnarray}
M_c = \frac{\alpha_s}{3} \sum_{i< j} < H_1| \lambda_i \cdot \lambda_j
\vec \sigma_i \cdot \vec \sigma_j \int r dr r' dr' \frac{g_i(r)f_i(r)}
{\chi^4(r)} \frac{g_j(r')f_j(r')}{\chi^4(r')} G_1(r,r')|H_1>
\end{eqnarray}     
where $g$ and $f$ are the upper and lower components of Dirac spinor,
$\chi^4$ is the colour dielectric function and
the Green's function $G_1(r,r')$ is defined in ref\c{9,Willets}.

The meson self energy $M_{meson}$ is computed by coupling the pseudoscalar
meson octate to the dihyperon. The meson coupling to the dihyperons leads
to an octate of dibaryon states and the meson energy calculation requires
the wavefunctions and masses of these states. The octate dibaryon wavefunctions
are obtained by operating $\sum_i \lambda^a_i \vec \sigma_i$ on the dihyperon 
state. Here $\lambda^a_i$ are the ( flavour ) SU(3) Gell-Mann matrices for 
$i^{th}$ quark and $\vec \sigma_i$ is the spin operator.  
The masses of the dibaryon octate have been calculated by using the procedure 
outlined in the preceeding paragraphs. 

The expression for the meson self energy is
\begin{eqnarray}
M_{meson} &=& \frac{3}{2 f_\pi^2 \pi^2} {\Big [} 3 \int \frac{ k^4 dk \;\; 
v^2_\pi (k)
}{\epsilon_\pi(k)(M_0 - M_\Sigma -\epsilon_\pi(k))} \nonumber \\
 & & +  2 \int \frac{ k^4 dk v^2_K (k)} {\epsilon_K(k)(M_0 -M_N -\epsilon_K(k))}
+ 2 \int \frac{ k^4 dk v^2_K (k)} {\epsilon_K(k)(M_0 -M_\Xi -\epsilon_K(k))}
\nonumber \\
& &  \int \frac{k^4 dk v^2_\eta (k)} {\epsilon_\eta(k)(M_0 -M_\Lambda 
- \epsilon_\eta(k))} ]
\end{eqnarray}
Here $M$'s are the masses of the dibaryon states excluding the meson self 
energy,
{\footnote {The notation used for the dibaryon octate is same as 
that of the baryon octate. Thus, $M_N$ is the mass of the dibaryon with isospin
1/2, hypercharge 1 and spin 1. It should not be confused with the nucleon mass}}
 $v(k)$ is the form factor for the meson coupling to the 
quark in the dibaryon states and $\epsilon(k)$ is the energy of the 
respective meson. 
Since the dihyperon state is a spin-flavour singlet, 
the pseudoscalar meson octate induces a transition between
the dihyperon and  the ( flavour ) octate of the dibaryon states having 
spin 1. The dibaryon octate states, in turn, couple to other dibaryon
states through the coupling of the pseudoscalar meson octate.
Thus, in a complete calculation, the coupling of the octate dibaryons with
other dibaryon states should be included and the masses of all the dibaryon 
states should be determined in a consistant fashion. 
Such a somewhat formidable calculation
is being done. Here we want to present the results of a restricted calculation
described above.

We now come to the discussion of the results of our calculation.
The parameters of the  CCD model are the quark masses( $m_0$ and $m$),
the strong coupling constant $\alpha_{s}$, the bag pressure $B$, the 
meson-quark coupling constant $f$, the mass of the dielectric field $M_{GB}$ 
and the constant $\alpha$. Of these, we have fixed $f = 93 MeV$ ( the pion
decay constant ) and $\alpha = 24$. Rest of the parameters have been chosen
by fitting the octate and decuplet baryon masses. Earlier\c{9} we have shown 
that the fitting procedure yields  a limited range of values of $m_0$,
$B$ and $m$ ( $m_0 \sim 100 MeV$, $m \sim 200 MeV$,
and $100 MeV < B^{1/4} < 140 MeV$) where as the strong coupling constant 
$\alpha_s$ is essentially determined by nucleon-delta mass difference. 
But equally good fits to the baryon masses are obtained for a wide
range of the glueball mass ( $0.8 GeV < m_{GB} < 1.5 GeV$ ). However, it has 
been found that the lower values of glueball mass yield better values of 
charge radii and magnetic moments. In the present calculation 
the dihyperon mass has been computed for a large set  of the parameter values 
which fit the masses of nucleon, $\Delta$ and $\Lambda$ masses. For 
these sets, the difference between calculated masses of other octate and 
decuplet baryons and the experimental masses is within few \%. Moreover, 
the variation in the individual baryon  masses over the range of the parameter
set considered here is only 10-15 MeV.

\begin{table}
\begin{center}
\caption{The dependence of dihyperon mass ($M_0$) on the parameters of
the CCD model. $\alpha_s$ is dimensionless and the masses are in the units 
units of MeV.} 

\vspace{1.0cm}
\begin{tabular}{|rcccc|c|}
\hline
\hspace{1.5cm} &\hspace{1.5cm} &\hspace{1.5cm} &\hspace{1.5cm} 
&\hspace{1.5cm} &\hspace{1.5cm}  \\
$m_{GB}$& $\alpha_s$ & $m_0$ & $m$ & $b^{1/4}$ & $M_0$  \\
 & & & & &  \\
\hline
   819. &   .269 &   103. &   211. &   103. &   2071. \\ 
   893. &   .474 &   123. &   210. &   104. &   2073. \\ 
   928. &   .288 &   108. &   212. &   108. &   2087. \\ 
   967. &   .581 &   133. &   209. &   106. &   2083. \\ 
  1008. &   .216 &   102. &   213. &   113. &   2103. \\ 
  1059. &   .272 &   111. &   213. &   115. &   2109. \\ 
  1118. &   .260 &   112. &   213. &   118. &   2119. \\ 
  1168. &   .436 &   132. &   211. &   118. &   2121. \\ 
  1207. &   .232 &   112. &   214. &   123. &   2133. \\ 
  1251. &   .215 &   111. &   215. &   125. &   2140. \\ 
 & & & & &  \\
\hline
\end{tabular}
\end{center}
\end{table}

The calculated dihyperon masses are displayed in Table 1. In the results
shown here the glueball mass
has been varied from 800 MeV to 1250 MeV, the u and d quark mass is varied 
between 100 and 135 MeV and $B^{1/4}$ is varied between 100 and 125 MeV.  
It is interesting to note that the
dihyperon mass increases  almost linearly with the glueball mass and does not 
show any systematic dependence on the other parameters. 
Further more, the variation in the dihyperon mass is quite large. For
example, the dihyperon mass changes by about 
70 MeV when the glueball mass is increased from about 800 MeV to 1250 MeV. This 
variation in the dibaryon mass arising from the chenge in $m_{GB}$ 
is about an  order of magnitude larger than the variation found in
the baryon octate and decuplet masses. Here we would like to note that the 
lower values of the glueball masses ( $m_{GB} < 1 GeV$ ) yield better 
agreement with the static properties of baryons ( charge radii, magnetic 
moments etc )\c{9}. Further more, it has been observed that a better agreement
with the $\pi N$ scattering data is obtained for $m_{GB} \sim 1 GeV$ or smaller.
We therefore feel that results with $m_{GB} \leq 1 GeV $ are somewhat more
realistic.

The results in Table 1 show that the computed dihyperon masses are smaller 
than the $\Lambda-\Lambda$ threshold. Thus, the dihyperon is stable against 
strong decays in the CCD model. The binding energy of the dihyperon varies 
between $160 MeV$ ( for $m_{GB}$ of $800 MeV$ ) and $90 MeV$ ( for $m_{GB}$
of $1250 MeV$ ). These values are larger than the result of S. Ahmed et al.
\c{Long} as well as Jaffe's prediction\c{jaf77}. 

To conclude, we have calculated the dihyperon mass using CCD model. Along with
the colour magnetic energy, we have also investigated the effect of the 
quark-meson coupling on the dihyperon mass. The correction due to the spurious
motion of the center of mass is included in the calculation. The projection
technique is used to project out the good momentum states and these states are
used in the computation of the dibaryon-meson form factors. It is found that
the dihyperon is  
stable against the strong decays for the parameters of the CCD model 
considered in the calculations with the binding energy of about 100 MeV or 
more.
The determination of the dihyperon width ( due to the weak interaction ), 
masses of other dibaryons and 
their strong decay widths ( due to their decay into a pair of baryons ) in the 
CCD model needs to be done. These calculations are in progress.

Our results are significant in the context of ongoing search for the 
quark-gluon plasma in the laboratory. One of the possible unambiguous way
to detect the transient existence of a temporarily created QGP might be
the experimental observation of exotic remnants like the formation of 
strange matter or strangelets \c{baym,greiner}. The six-quark dihyperon state
is supposed to be the lightest strangelet state. So the fact that such states 
are found to be stable for the reasonable parameter ranges in the present 
study, makes it imperative to put more experimental efforts to detect such 
objects.
 
SKG would like to thank Council for Scientific and Industrial Research (CSIR),
 Govt. of India for financial support.
 
\end{document}